# L(2,1)-labelling of Circular-arc Graph

*Satyabrata Paul*[1]*, Madhumangal Pal*[1] *and Anita Pal*[2]

[1]Department of Applied Mathematics with Oceanology and Computer Programming, Vidyasagar University, Midnapore-721102, India
e-mail: {satyamath09,mmpalvu}@gmail.com

[2]Department of Mathematics
National Institute of Technology Durgapur, Durgapur-713209, India
e-mail: anita.buie@gmail.com

**Abstract.** An $L(2,1)$-labelling of a graph $G=(V,E)$ is $\lambda_{2,1}(G)$ a function $f$ from the vertex set $V(G)$ to the set of non-negative integers such that adjacent vertices get numbers at least two apart, and vertices at distance two get distinct numbers. The $L(2,1)$-labelling number denoted by $\lambda_{2,1}(G)$ of $G$ is the minimum range of labels over all such labelling. In this article, it is shown that, for a circular-arc graph $G$, the upper bound of $\lambda_{2,1}(G)$ is $\Delta+3\omega$, where $\Delta$ and $\omega$ represents the maximum degree of the vertices and size of maximum clique respectively.

**1. Introduction**
The frequency assignment problem is to assign frequency to a given group of televisions or radio transmitters so that interfering transmitters are assigned frequency with at least a minimum allowed separation. This problem was formulated as a vertex colouring of graph by Hael [21]. In 1988, Roberts proposed a variation of the frequency assignment problem in which `close' transmitters must receive different channels and `very close' transmitters must receive channels at least two apart. To convert this problem into graph theory, the transmitters are represented by the vertices of a graph; two vertices $x$ and $y$ are `very close' if the distance between them is 1 and `close' if the distance between $x$ and $y$ is 2. We denote, $d(x,y)$ to represent the shortest distance (i.e. the minimum number of edges) between the vertices $x$ and $y$. The definition of $L(2,1)$-labelling problem of a graph is given below.

**Definition 1.** $L(2,1)$-labelling of a graph $G=(V,E)$ is a function $f$ from $V$ to the set of non-negative integers $\{0,1,2,...,\lambda\}$ such that $|f(x)-f(y)|\geq 2$ if $d(x,y)=1$ and $|f(x) - f(y)| \geq 1$ if $d(x, y) = 2$, where $\lambda$ is a suitable integer.

The span of $L(2,1)$-labelling $f$ of $G$ is the difference between largest and smallest used labels. The minimum span over all possible labelling functions is denoted by $\lambda_{2,1}(G)$ or simply $\lambda(G)$.

Many results have been published related to this problem and its variations [10, 55]. The determination of exact value of $\lambda_{2,1}(G)$ for a given graph $G$ is a very difficult task. For this reason, people are trying to determine the upper bounds for different classes of graphs. An obvious lower bound is $\Delta + 1$, where $\Delta = \max\{\deg(v) : v \in V\}$ is the maximum degree of the graph. Griggs and Yeh [18] first shown that for any graph $G$, $\lambda_{2,1}(G) \leq \Delta^2 + 2\Delta$. The above bound was improved to by Chang and Kuo [13]. Kral' and Skrekovski [30] proved that $\lambda_{2,1}(G) \leq \Delta^2 + \Delta - 1$. It was again improved to $\lambda_{2,1}(G) \leq \Delta^2 + \Delta - 2$ by Goncalves [17]. Griggs and Yeh [18] conjectured that for any graph $G$, $\lambda_{2,1}(G) \leq \Delta^2$. Havet et al. [20] have proved this result asymptotically. Although this remains an open problem. But, this conjecture is true for several graph classes, such as paths, cycles, wheels [18], trees [13, 19], cographs [14], generalized petersen graphs [24], chordal graphs [54], weekly chordal graphs [12], dually chordal graphs [45], strongly orderable graphs [45], outerplanar graphs [9], permutation graphs [6, 46], bipartite permutation graphs [1], interval graphs [5, 13, 47, 48], cactus graphs [25, 26, 27], regular grids [4, 11], block graphs [7], direct product of triangle and a cycles [28], $K_n \times K_2$ [29], etc. There are very few graph classes for which $\lambda_{2,1}(G)$ can be calculated efficiently. These are paths, cycles, wheels, trees, cactus graphs, cographs, regular grids, generalized petersen graphs, etc. There are some graph classes for which it is still unknown that the computation of $\lambda_{2,1}(G)$ is NP-complete or polynomially solvable. Such classes of graphs are interval graphs, circular-arc graphs [8], chordal graphs [54], weekly chordal graphs [12], dually chordal graphs [45], strongly orderable graphs [45], outerplanar graphs [9], bipartite graphs, planar graphs, split graphs, permutation graphs [6, 46], bipartite permutation graphs [1], block graphs [7], hypercubes [13], etc. Thus finding good upper bounds of these graphs is a good result.

Circular-arc graph is a very important subclass of intersection graphs and has been widely studied in the past [34, 35, 36, 50]. Many works have been done on other intersections graphs, see[2,3,15,22,23,31-33,38-44,49,51-53]. Calamoneri et al. [8] shown that for circular-arc graphs $\lambda_{h,k}(G) \leq \max(h, 2k)\Delta + h\omega$. Motivated from this we study on circular-arc graphs and a good upper bound of these graphs and this result is more better than the previous upper bound [8].

The remaining part of the paper is organized as follows. Some properties of circular-arc graph are discussed in the next section. Section 3 contains the proof of main theorem. Finally, in Section 4 conclusions are made.

## 2. Preliminaries
The graphs used in this work are simple, without self loop or multiple edges. Let

## L(2,1)-labelling of Circular-arc Graph

$G = (V, E)$ be a graph with set of vertices $V$ and set of edges $E$.

A set $C \subseteq V$ is called a clique if for every pair of vertices of $C$ has an edge. The number of vertices of the clique represents its size. A clique $C$ of a graph $G$ is called maximal if there is no clique of $G$ which properly contains $C$ as a subset. Again, a clique with r vertices is called $r$-clique. A clique is called maximum if there is no clique of $G$ of larger cardinality. The number of vertices of the maximum clique of $G$ is denoted by $\omega(G)$ and is called a clique number of $G$. The neighbourhood of a vertex $x$ is $N_G(x) = \{y \mid xy \in E\}$. The set $N_G(x)$ is also known as 1-nbd vertex set of $x$ or $N_1(x)$. Similarly, 2-nbd vertices of $x$ is denoted by $N_2(x)$) and is defined as $N_2(x) = \{y \mid d(x, y) = 2\}$. The maximum degree of a graph $G$, denoted by $\Delta(G)$, is the maximum of the degrees of all the vertices. Throughout this paper we use $\Delta$ instead of $\Delta(G)$.

**Definition 2.** (Circular-arc graph) An undirected graph $G = (V, E)$ is said to be a circular-arc graph if there exists a family $A$ of arcs around a circle and a one-to-one correspondence between vertices of $G$ and arcs in $A$, such that two distinct vertices are adjacent in $G$ if and only if the corresponding arcs intersects in $A$. Such a family of arcs is called an arc representation for $G$.

It is assumed that all the arcs must cover the circle, otherwise the circular-arc graph is nothing but an interval graph.

Let $A = \{A_1, A_2, \ldots, A_n\}$ be a set of arcs around a circle. While going in a clockwise direction, the point at which we first encounter an arc will be called the *starting point* of the arc. Similarly, the point at which we leave an arc will be called the *finishing point* of that arc. An arc $A_i$ is denoted by a closed interval $[s_i, f_i], i = 1, 2, \ldots, n$, where, $s_i$ is the counter clockwise end i.e., starting point and $f_i$ is the clockwise end i.e., finishing point and obviously $s_i < f_i$. To illustrate our problem we consider a circular-arc graph of Figure 1.

We consider a fixed line drawn from the center of the circle and passing through the finishing point of any arc. Some arcs are intersected by this line. This set of arcs is called the set of *backward arcs* and the remaining sets of arcs are called the set of *forward arcs*. If any two arcs $A_i$ and $A_j$ share the common points of the circle the arcs are said to be *intersecting arcs*. If two arcs do not share a

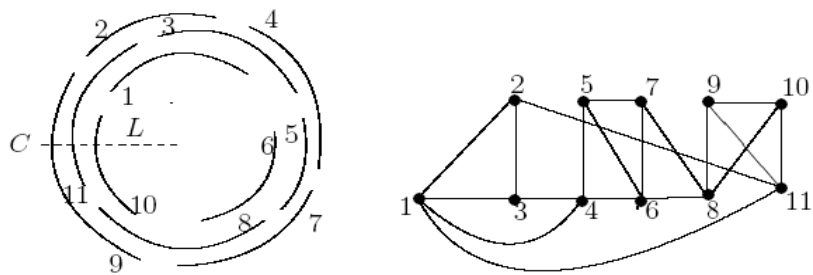

Figure 1. A circular-arc representation and the corresponding circular-arc graph G.

Common point of the circle then they are called *independent arcs*.

Without loss of generality, we assume the following:

(i) No two arcs share a common endpoint.

(ii) The arcs are sorted in increasing values of $s_i$ i.e. $s_i > s_j$ for $i > j$.

Now, the adjacency between two arcs (vertices) $A_i = [s_i, f_i]$ and $A_j = [s_j, f_j]$ can be tasted by using the following lemma.

**Lemma 1.** If two arcs $A_i = [s_i, f_i]$ and $A_j = [s_j, f_j]$ are adjacent then one of the following conditions is true

(i) $s_i < s_j < f_i < f_j$ or

(ii) $s_i < s_j < f_j < f_i$ or

(iii) $s_j < s_i < f_i < f_j$.

**Proof.** Let $s_i < s_j < f_i < f_j$. This implies that the starting point of arc $A_j$ lies between starting point and finishing point of arc $A_i$. Then the portion $[s_j, f_i]$ is common for both arcs $A_i$ and $A_j$. So the intersection of these two arcs is non-empty. Hence $A_i$ and $A_j$ are adjacent.

If $s_i < s_j < f_j < f_i$, it means the starting and finishing points of arc $A_j$ lies between starting and finishing points of $A_i$. Then arc $A_j$ is properly contained by $A_i$. So their intersection is non-empty. Hence arcs $A_i$ and $A_j$ are adjacent.

Similarly, let $s_j < s_i < f_i < f_j$. In this case the arc $A_i$ is dominated by $A_j$. The portion $[s_i, f_i]$ is common for both the arcs. So their intersection is non-empty. Hence arcs $A_i$ and $A_j$ are adjacent. □

## 3. The $L(2,1)$-labeling of circular-arc graph

Calamoneri et al. [8] have shown that for a circular-arc graph $G$, $\lambda_{h,k}(G) \leq \max(h, 2k)\Delta + h\omega$. Therefore, for $h = 2$ and $k = 1$, the above result becomes $\lambda_{2,1}(G) \leq 2\Delta + 2\omega$. In this paper, we present a new method for $L(2,1)$-labelling of circular-arc graph and shown that the new upper bound for $\lambda_{2,1}$ is $\Delta + 3\omega$. Clearly, $\Delta \geq \omega$ for any graph. Thus, this bound is more tight than the previous one.

Any circular-arc graph can be transferred to an interval graph by deleting some appropriate arcs. The main idea to $L(2,1)$-label a circular-arc graph is to convert the given graph to an interval graph by deleting some suitable arcs. In [48], Paul et al. have designed an algorithm for $L(2,1)$-labelling of an interval graph. They have established the following result.

**Theorem 1.** For any interval graph $G$, $\lambda_{2,1}(G) \leq \Delta + \omega$.

## $L(2,1)$-labelling of Circular-arc Graph

To convert a circular-arc graph to an interval graph draw a line $L$ from the center of the circle perpendicular to the arcs (see Fig. 1). Let $C$ be the set of arcs which intersects the line $L$. Remove the arcs of $C$ from the set of arcs $A$. If we consider the set of arcs $A - C$ as line segments, then they forms a set of intervals on a real line (Fig. 2) and these intervals form an interval graph. Thus we state the following result.

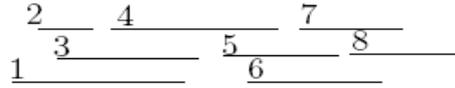

Figure 2. The set of intervals obtained from Fig. 1 by deleting the arcs of $C$.

**Lemma 2.** The arcs of $A - C$ induced an interval representation on a real line.

The set of arcs $C$ satisfy the following result.

**Lemma 3.** The vertices corresponding to the arcs of $C$ form a clique.
**Proof.** The arcs of $C$ must intersect the line which is drawn from the center of circle. That is, any two arcs intersect each other. Thus the vertices corresponding to the arcs of $C$ forms a clique. □

The main idea for $L(2,1)$-labelling of interval graph presented in [48] is discuss below.

Let $V = \{v_1, v_2, \ldots, v_n\}$ be the set of vertices such that $v_1 < v_2 < \ldots < v_n$. We choose the first maximal clique (i.e. the clique contains the vertex $v_1$). Let $C_1$ be the first maximal clique and arrange the vertices of $C_1$ according to the decreasing order of the degree of the vertices and let this set be $C_{1'}$. Label the vertices of $C_{1'}$ by $0, 2, \ldots, 2(p-1)$, where $p$ is the cardinality of the set $C_1$. That is, $C_1 = \{v_1, v_2, \ldots, v_p\}$. Now we compute the set $V - C_1$ and label each vertex by first available label such that the condition of $L(2,1)$-labelling for $G$ is satisfied.

We define some sets as follows.
$A_i = \{v_j : d(v_j, v_i) = 1 \text{ and } j < i\}$ and
$B_i = \{v_j : d(v_j, v_i) = 2 \text{ and } j < i\}$,
i.e., $A_i$ is the set of 1-nbd labelled vertices of $v_i$ and $B_i$ is the set of 2-nbd labelled vertices of $v_i$. Clearly, $A_i \subseteq N_1(i)$ and $B_i \subseteq N_2(i)$.

Moreover, the indices of the vertices of the sets $A_i$ and $B_i$ are strictly less then $i$. Now, by $L(2,1)$-labelling, $f_i$, the label of $v_i$ depends only on the label of each vertices of the sets $A_i$ and $B_i$ and not any other vertices. Again, we define two sets $r_{N_1(v_i)}(f_j)$ and $r'_{N_2(v_i)}(f_j)$ as follows:

$r_{N_1(v_i)}(f_j)$ is the set of integers forbid by the label $f_j$, where $v_j \in N_1(v_i)$ and

$r'_{N_2(v_i)}(f_j)$ is the set of integers forbid by the label $f_j$, where $v_j \in N_2(v_i)$. That is, $r'_{N_1(v_i)}(f_j) = \{f_j - 1, f_j, f_j + 1\} \cap (Z^+ \cup \{0\})$ where $v_j \in N_1(v_i)$, and $r'_{N_2(v_i)}(f_j) = \{f_j\}$.

Again, $S_i^1$ is the set of integers forbid by all the vertices of $A_i$ and $S_i^2$ is the set of integers forbid by all the vertices of $B_i$. We define $R_i$ as $R_i = S_i^1 \cup S_i^2$.

**Algorithm L21CA**
**Input:** Circular-arc representation of the given circular-arc graph $G$.
**Output:** $L(2,1)$-labelling of $G = (V, E)$.

**Step 1:** Draw a straight line $L$ and let $C$ be the set of arcs intersecting $L$.

**Step 2:** Remove $C$ from the circular-arc representation. Let this reduced graph be $G' = (V', E')$, where $V'$ is the set of vertices corresponding to the arcs $A - C$.

**Step 3:** Compute the first maximal clique $C_1$ of $G'$.

**Step 4:** Arrange the vertices of the clique $C_1$ according to the decreasing order of the degree of the vertices. If the degree of two vertices $v_i, v_j \in C_1$ are equal then we put $v_j$ before $v_i$ if $i > j$, in the order and let this new clique be $C_1'$.

**Step 5:** // Label all the vertices of $C_1'$ //

for each vertex $v_k \in C_1'$

set $f_k = 2(i-1)$ // $i$ represents the position of the vertex $v_k$ in $C_1'$ //

**Step 6:** Compute the set $V' - C_1$.

If $V' - C_1 = \phi$ then
Stop.
Otherwise go to Step 5.

**Step 7:** for $i = m$ to $n$ do // $v_m$ is the first vertex of $V' - C_1$ //

Compute $A_i = \{v_j : d(v_j, v_i) = 1 \text{ and } j < i\}$;

and $B_i = \{v_j : d(v_j, v_i) = 2 \text{ and } j < i\}$.

**Step 8:** Compute for each vertex $v_j \in A_i$

$$r_{A_i}(f_j) = \{f_j - 1, f_j, f_j + 1\} \cap (Z^+ \cup \{0\})$$

and for each vertex $v_j \in B_i$

$$r'_{B_i}(f_j) = \{f_j\}.$$

**Step 9:** $S_i^1 = \bigcup_{v_j \in A_i} (r_{A_i}(f_j))$, //set of positive integers forbid by all the vertices of $A_i$//

$S_i^2 = \bigcup_{v_j \in B_i} (r'_{B_i}(f_j))$. //set of positive integers forbid by all the vertices of $B_i$//

**Step 10:** $R_i = S_i^1 \cup S_i^2$ // $R_i$ is the set of positive integers unavailable for $v_i$ //

**Step 11:** $f_i = \min\{(Z^+ \cup \{0\}) - R_i\}$
end for;
Let $F$ be the highest label, i.e. maximum of $f_i$.

**Step 12**: Now we label the vertices corresponding to the arc of $C = \{A_{c_1}, A_{c_2}, \ldots, A_{c_k}\}$.
for $j = 1$ to $k$
$f_{c_j} = F + 2 + 2(j-1)$; // $f_{c_j}$ is the label of the vertex corresponding to the arc $A_{c_j} \in C$
.//
end for;
**end L21CA.**

### 3.1. Illustration

We now illustrated this algorithm by a suitable example. We choose the graph of Fig. 1 and label it by $L(2,1)$-labeling. For this graph, $\Delta = 4$ and $\omega = 3$.

We draw a line to get $C$. In Fig 1. $C = \{9, 10, 11\}$.

Now, by Step 2, we remove $C$ from the circular-arc representation and the remaining graph is an interval graph $G' = (V', E')$, where $V' = \{v_1, v_2, v_3, \ldots, v_8\}$ corresponding to the arc $1, 2, 3, \ldots, 8$.

In Step 3, $C_1 = \{v_1, v_2, v_3\}$ is the first maximal clique of $G'$. Here $C' = C = \{v_1, v_2, v_3\}$.

Now we label the vertices of $C'$ by 0, 2, 4. So, $f_1 = 0, f_2 = 2, f_3 = 4$.

In Step 6, we compute $V' - C_1$ i.e. $\{v_4, v_5, v_6, v_7, v_8\}$. So, $V' - C_1 \neq \phi$. Now we want to label the vertex $v_4$.

$A_4 = \{v_1, v_3\}$ and $B_4 = \{v_2\}$.
Therefore,
$r_{A_4}(f_1) = \{-1, 0, 1\} \cap (Z^+ \cup \{0\})$
$= \{0, 1\}$
and $r'_{B_4}(f_2) = \{2\}$.
Similarly,
$r_{A_4}(f_3) = \{3, 4, 5\} \cap (Z^+ \cup \{0\})$
$= \{3, 4, 5\}$.
So, $S_4^1 = r_{A_4}(f_1) \cup r_{A_4}(f_3) = \{0, 1, 3, 4, 5\}$ and $S_4^2 = \{2\}$.
Thus, $R_4 = S_4^1 \cup S_4^2 = \{0, 1, 2, 3, 4, 5\}$.
$\therefore f_4 = \min\{(Z^+ \cup \{0\}) - R_4\}$
$= 6$.
Similarly, $f_5 = 1, f_6 = 3, f_7 = 5, f_8 = 0$.

Here, $F = 6$.
By Step 12, we label the vertices of $C$. Here $C = \{9, 10, 11\}$.
$$\therefore f_9 = F + 2 + 2(1-1)$$
$$= 6 + 2 = 8.$$
Similarly, $f_{10} = 6 + 2 + 2 \times 1 = 10$ and $f_{11} = 12$.

### 3.2. Analysis of algorithm
Following theorem gives the correctness of the proposed algorithm.

**Theorem 2.** Algorithm L21CA correctly labels a circular-arc graph maintaining $L(2,1)$-labelling conditions.

**Proof.** In algorithm L21CA, first we label the set $C_1 \in G'$, i.e. the first maximal clique. We use the labels $0, 2, 4, \ldots, 2(p-1)$, where $p$ is the cardinality of the set $C_1$. So, it follows $L(2,1)$-labelling conditions. Now we label the set $V' - C_1$. Since our labelling procedure is $L(2,1)$-labelling, so when we proposed to label a vertex $v_i$, then its label depend only on the vertices of 1-nbd and 2-nbd of $v_i$, not any other vertices. Thats why we define the sets $A_i$ and $B_i$.

For each vertex $v_j \in A_i$, the forbidden integers by $f_j$ are $f_j - 1, f_j$, and $f_j + 1$ as $d(v_i, v_j) = 1$. Again for each $v_l \in B_i$, the forbidden integers by $f_l$ is $f_l$ only as $d(v_l, v_i) = 2$. In Step 10, we compute all unavailable labels for $v_i$ and remove these integers from $Z^+ \cup \{0\}$. Thus the remaining set of integers is available for $v_i$ and we choose the minimum one. So, each vertex $v_i \in V' - C_1$ follows $L(2,1)$-labelling conditions. Next we label the vertices corresponding to the arcs of $C$ by $0, 2, 4, \ldots, 2(k-1)$ ($k$ is the cardinality of the set $C$) as $C$ form a clique (by lemma 3). This additional label start from $F + 2$, where $F$ is the maximum label used to label the vertices of $V' - C_1$ and 2 is added to maintaining the labelling conditions. So algorithm L21CA correctly label the whole graph $G$ as $G = C_1 \cup (V' - C_1) \cup C$. □

Using the above algorithm, one can $L(2,1)$-label the vertices of a circular-arc graph, and the number of labels needed is $\Delta + 3\omega$, justified below. But, $\Delta + 3\omega$ is not necessary the optimum (i.e. minimum) number of labels for a given circular-arc graph. That graph may be label by less number of labels.

**Theorem 4.** For any circular-arc graph $G = (V, E)$, $\lambda_{2,1}(G) \leq \Delta + 3\omega$, where $\Delta$ and $\omega$ represent the maximum degree of the vertices and size of maximum clique of $G$ respectively.

**Proof.** Let $G'$ is an interval graph corresponding to the intervals representation of $A - C$ (by Lemma 2). $\Delta_1$ and $\omega_1$ represents the degree of the vertices and size of maximum

*L*(2,1)-labelling of Circular-arc Graph

clique of $G'$ respectively. Clearly, $\Delta_1 \leq \Delta$ and $\omega_1 \leq \omega$ as $G' \subseteq G$. Thus, by Theorem 1, $\lambda_{2,1}(G') \leq \Delta_1 + \omega_1$. Again, the vertices of $C$ forms a clique (by Lemma 3), and it can be labelled by the integers $0, 2, 4, \ldots, 2(|C|-1)$. Thus, to label all the vertices corresponding to the arcs of $C$, at most $2(\omega-1)+1$ additional labels are required as $\max|C| = \omega$. These additional labels must be started from $\Delta_1 + \omega_1 + 2$ as from Theorem 1, $\lambda_{2,1}(G') \leq \Delta_1 + \omega_1$.

Therefore,
$\lambda_{2,1}(G) \leq \Delta_1 + \omega_1 + 2 + 2(\omega-1).$

Thus, $\lambda_{2,1}(G) \leq \Delta + 3\omega$ ($\because \Delta_1 \leq \Delta$ and $\omega_1 \leq \omega$).

Hence the result. □

**Theorem 3.** Time complexity of algorithm L21CA is $O(n^2)$, where $n$ is the number of vertices of the graph.

**Proof.** The set of arcs $C$ can be computed in $O(n)$ time. Again, to remove all these arcs from $A$ takes $O(n)$ time. After removing $C$ from $A$, the resulting graph $G' = (V', E')$ becomes an interval graph, where the vertices of $V'$ corresponds to the set of arcs $A - C$. The first maximal clique (Step 3) of an interval graph can be computed in $O(n)$ time [37]. Steps 4, 5 and 6 are straight forward and these steps take only $O(n)$ time. Step 7 computes $A_i$ and $B_i$ for $n$ vertices. Each set takes $O(n)$ times. Thus, total time for step 7 is $O(n^2)$. The remaining steps take $O(n)$ time. Hence the overall time complexity of the algorithm L21CA is $O(n^2)$. □

### 4. Concluding remarks
In this paper, we find a new upper bound of $L(2,1)$-labeling of circular-arc graphs. These bounds are more tight than the previous available results $2\Delta + 2\omega$ [8]. But, we are not able to find exact value of $\lambda_{2,1}(G)$ for a given circular-arc graph. We feel that it is very difficult to determine the exact value of $\lambda_{2,1}(G)$ for a given circular-arc graph. Complexity of this problem is still open.

**Acknowledgement**
Financial support offered by the Department of Science and Technology, New Delhi, India (Ref. No.SR/S4/MS/655/10) is thankfully acknowledged.